\documentstyle[prb,aps,twocolumn,epsf]{revtex}

\begin{document}

\twocolumn[\hsize\textwidth\columnwidth\hsize\csname
@twocolumnfalse\endcsname

\title{Phase diagram of the one-dimensional extended attractive Hubbard model for
large nearest-neighbor repulsion}
\author{A. A. Aligia}
\address{Comisi\'on Nacional de Energ{\'\i}a At\'omica,\\
Centro At\'{o}mico Bariloche and Instituto Balseiro, 8400 S.C. de Bariloche,%
\\
Argentina.}
\date{Received \today }
\maketitle

\begin{abstract}
We consider the extended Hubbard model with attractive on-site interaction $%
U $ and nearest-neighbor repulsions $V$. We construct an effective
Hamiltonian $H_{eff}$ for hopping $t<<V$ and arbitrary $U<0$. Retaining the
most important terms, $H_{eff}$ can be mapped onto two $XXZ$ models, solved
by the Bethe ansatz. The quantum phase diagram shows two Luttinger liquid
phases and a region of phase separation between them. For density $n<0.422$
and $U<-4$, singlet superconducting correlations dominate at large
distances. For some parameters, the results are in qualitative agreement
with experiments in Ba$_{1-x}$K$_{x}$BiO$_{3}$.
\end{abstract}


\vskip2pc]

\narrowtext

\section{Introduction}

BaPb$_{1-x}$Bi$_{x}$O$_{3}$ and Ba$_{1-x}$K$_{x}$BiO$_{3}$ are
superconductors with transition temperatures $T_{c}$ near 13K \cite{slei}
and 30K \cite{matt} respectively. In spite of the fact that these systems do
not contain Cu and are non-magnetic \cite{batt}, they have important
similarities with the cuprate superconductors, like the perovskite
structure, relatively high $T_{c}$ and low density of states at the Fermi
level \cite{batt}. While the electron-phonon interaction is very important
in these materials, as shown by inelastic neutron measurements \cite{loon},
and the existence of displacements of O atoms and lattice distortions \cite
{cox,chai,pei}, several other experiments suggest that the pairing mechanism
is, at least in part, of electronic nature \cite{batt}. Excitons combined
with O displacements \cite{dolo,sofo}, and on-site attractive interaction $U$
based on non-linear screening \cite{varm,nguy}, have been proposed as the
origin of superconductivity in doped BaBiO$_{3}$. In these approaches,
repulsive interactions at finite distances play an important role, what is
consistent with poor screening of Coulomb interactions by the low density of
carriers.

As first shown by Rice and Sneddon \cite{rice}, treating the displacements $%
d $ of O ions in the direction of their nearest-neighbor Bi ions in the
antiadiabatic approximation, leads to a decrease in $U$ by $zg^{2}/K$, and
an increase of the repulsion $V$ between nearest Bi atoms by $g^{2}/K$,
where $z$ is the coordination number, $g$ the electron-phonon interaction,
and $K$ the second derivative of the elastic energy with respect to $d$.
This leads naturally to the extended Hubbard model for the description of
doped BaBiO$_{3}$, and to superconductivity carried by bipolarons (a pair of
carriers at Bi sites accompanied by O displacements) \cite{rice,micn}. While
the excitonic mechanism involves partial occupation of O states by holes 
\cite{dolo}, in agreement with optical experiments \cite{sale,heal,uchi}, it
is still possible that this one-band model describes the low-energy physics
(as in the cuprates \cite{sim}). This is certainly the case in the purely
electronic model proposed by Varma, in which non-linear screening not only
reduces the bare atomic $U\sim 11$ eV, but renders it negative \cite{varm}.

In recent years, the phase diagram of the extended Hubbard model in one
dimension has been studied by numerical techniques, with particular emphasis
in the quarter filled case (number of particles per site $n=1/2$) \cite
{mila,penc,sano,lin,ali,clay,nak}. For $n=1/2$ or $n=2/3$, $V>5$ and small
negative $U$, calculations of the correlation exponent $K_{\rho }$ in
systems of up to $L=16$ sites, predict a Luttinger liquid phase with
dominant superconducting correlations at large distances \cite{penc,sano}.
However, from an analysis based on the infinite $V$ limit and the
extrapolation of the spin gap, Penc and Mila suggested that the results of $%
K_{\rho }$ were affected by finite-size effects, and the system would not be
a Luttinger liquid in that region \cite{penc}. For $n=1/2$, perturbative
arguments around the infinite $V$ limit (developed in more detail here)
suggested that this region corresponds to phase separation \cite{ali}. Monte
Carlo studies in systems with $L\sim 64$ lattice sites \cite{clay} have
shown that the region of phase separation in the model extends to much lower
values of $V$ than those obtained for $L\leq 16$, leaving practically no
place for a Luttinger liquid phase with dominant superconducting
correlations for large $V$. The extent of this phase remains unclear.
Furthermore, due to the above mentioned finite-size effects, and technical
problems with different Monte Carlo methods \cite{clay}, the region of small 
$|U|$ and $V>8t$, where $t$ is the hopping, is practically unaccessible to
the present available numerical methods. This region is also out of the
range of applicability of the continuum-limit field theory (also called $g$%
-ology), which has been applied to the extended Hubbard and related models 
\cite{voitl,frad,voitb,bos}.

In this work, we study the one-dimensional attractive Hubbard model in the
limit $V>>t$. Extending previous work \cite{penc}, we derive an effective
low-energy Hamiltonian including terms up to second order in $t$. Under
certain approximations, which are not essential for sufficiently large $V$,
the effective Hamiltonian is mapped into two $XXZ$ models, representing the
movements of particles in singly and doubly occupied sites respectively.
From the Bethe ansatz solution of these models, the phase diagram in the
thermodynamic limit $L\rightarrow \infty $ as a function of $n$, $U/t<0$ and
large $V/t$ is obtained. In the next Section, we explain the model and the
mapping procedures. The results are shown in Section III. Section IV
contains the conclusions and a discussion of the possible relation between
our results and the phase diagram measured in Ba$_{1-x}$K$_{x}$BiO$_{3}$ 
\cite{pei}.

\section{Model and effective Hamiltonian}

The one-dimensional extended Hubbard Hamiltonian in standard notation is:

\begin{equation}
H=-t\sum_{i\sigma }(c_{i+1\sigma }^{\dagger }c_{i\sigma }+{\rm H.c.}%
)+U\sum_{i}n_{i\uparrow }n_{i\downarrow }+V\sum_{i}n_{i+1}n_{i},  \label{ehm}
\end{equation}
and we consider the case $U<0$, $V>>t$. In this limit, there are no
nearest-neighbor occupied sites in the low-energy subspace. The terms in Eq.
(\ref{ehm}) which mix states of this subspace with states with non-zero
occupancy at nearest-neighbor sites, can be eliminated through a standard
canonical transformation, as described in Ref. \cite{penc}. This procedure
originates terms of order $t^{2}/V$ or smaller in the low-energy subspace.
The resulting effective Hamiltonian within this subspace (a projector $%
P=\prod_{i\sigma \sigma ^{\prime }}(1-n_{i\sigma }n_{i+1\sigma ^{\prime }})$
is implicit) can be written in the form:

\begin{equation}
H_{eff}=H_{s}+H_{d}+H_{sd},  \label{h}
\end{equation}
where $H_{s}$ involves only singly occupied sites:

\begin{eqnarray}
H_{s} &=&-t\sum_{i\sigma }(c_{i+1\sigma }^{\dagger }c_{i\sigma }+{\rm H.c.}%
)(1-n_{i{\bar{\sigma}}})(1-n_{i+1{\bar{\sigma}}})  \nonumber \\
&&+V_{s}\sum_{i\sigma \sigma ^{\prime }}n_{i\sigma }n_{i+2\sigma ^{\prime
}}(1-n_{i{\bar{\sigma}}})(1-n_{i+2{\bar{\sigma}}})  \nonumber \\
&&+V_{s}\sum_{i\sigma \sigma ^{\prime }}(c_{i+3\sigma }^{\dagger
}c_{i+2\sigma }c_{i+1\sigma ^{\prime }}^{\dagger }c_{i\sigma ^{\prime }}+%
{\rm H.c.}),  \label{hs}
\end{eqnarray}
with $V_{s}=-t^{2}/V$. The second term is an interaction between second
nearest neighbors, and the third term displaces two next-nearest-neighbor
particles one lattice parameter.

The second term in Eq. (\ref{h}) can be described entirely in terms of the
operators $d_{i}^{\dagger }=c_{i\uparrow }^{\dagger }c_{i\downarrow
}^{\dagger }$, which create ``doublons'' at each site:

\begin{eqnarray}
H_{d} &=&\Delta _{d}\sum_{i}d_{i}^{\dagger }d_{i}-t_{d}\sum_{i\sigma
}(d_{i+1}^{\dagger }d_{i}+{\rm H.c.})  \nonumber \\
&&+V_{d}\sum_{i}d_{i+2}^{\dagger }d_{i+2}d_{i}^{\dagger }d_{i},  \label{hd}
\end{eqnarray}
where $\Delta _{d}=U-4t^{2}/(V-U)$, $t_{d}=2t^{2}/(V-U)$, and $%
V_{d}=4t^{2}[1/(V-U)-1/(3V-U)]$ are the effective on-site energy,
nearest-neighbor hopping, and next-nearest-neighbor repulsion respectively,
for doublons.

The last term in Eq. (\ref{h}) describes interactions between singly and
doubly occupied sites:

\begin{eqnarray}
H_{sd} &=&M_{1}\sum_{i}[(c_{i+1\uparrow }^{\dagger }c_{i-1\downarrow
}^{\dagger }+c_{i-1\uparrow }^{\dagger }c_{i+1\downarrow }^{\dagger }) 
\nonumber \\
&&\times (1-n_{i-1})(1-n_{i+1})(d_{i-1}+2d_{i}+d_{i+1})+{\rm H.c.}] 
\nonumber \\
&&+M_{2}\sum_{i\sigma }(c_{i\sigma }^{\dagger }d_{i+2}^{\dagger
}d_{i}c_{i+2\sigma }+{\rm H.c.})  \nonumber \\
&&+V_{sd}\sum_{i\sigma \delta =\pm 2}n_{i\sigma }(1-n_{i{\bar{\sigma}}%
})d_{i+\delta }^{\dagger }d_{i+\delta }.  \label{hsd}
\end{eqnarray}
Here $M_{1}=-t^{2}[1/V+1/(V-U)]/2$ describes annihilation of a doublon with
creation of two particles at empty sites, and the Hermitian conjugate
process, while $M_{2}=t^{2}/(2V-U)$ ($V_{sd}=t^{2}[2/(V-U)-2/(2V-U)-1/(2V]$)
corresponds to interchange (interaction) of a doublon and a particle at a
singly occupied next-nearest-neighbor site.

For $V=+\infty $, all terms of $H_{eff}$ vanish except the first term of Eq.
(\ref{hs}) and the first term of Eq. (\ref{hd}), and $H_{eff}$ can be solved
exactly \cite{penc}. In this limit, for $U>-4t$ and sufficiently small
density $n$, the system has no doubly occupied sites and is described by $%
H_{s}$. Instead, for $U<-4t$, $n\leq 1$, the ground state of $H_{eff}$ is
the same as that of $H_{d}$ and has no singly occupied sites. For other
values of $U$ and $n\leq 1$, the system phase separates into the phases just
described, and the limits of the region of phase separation (PS) can be
obtained using the Maxwell construction (finding the common tangent to the
curves $E_{s}(n)$ and $E_{d}(n)$, where $E_{\alpha }(n)$ is the ground state
of $H_{\alpha }$ at density $n$).

For the sake of clarity we call ``metallic'' (M) the phase without double
occupancy (ground state of $H_{s}$), and ``bipolaronic'' (BP), the phase
described by the ground state of $H_{d}$, although the negative $U$ is not
necessarily related with atomic displacements in a real system. For finite
but large $V$, the energy cost for constructing a uniform phase with both,
singly and doubly occupied sites is high in comparison with $M_{1}$, $M_{2}$
and $V_{sd}$. Thus, in the ground state, $H_{sd}$ can only act in the PS
region, at the boundary between M and BP phases, and is irrelevant in the
thermodynamic limit. Our main approximation is the neglect of $H_{sd}$. This
should be correct as long as the energy gain of $H_{s}$ in the PS region ($%
\sim t$) is larger than the terms of $H_{sd}$ ($\sim t^{2}/V$). We also
neglect the last term of Eq. (\ref{hs}). This term vanishes at the extreme
densities of the M phase ($n=0$ and $n=1/2$). The ratio of its expectation
value with respect to the expectation value of the first term of Eq. (\ref
{hs}) can be estimated by perturbation theory:

\begin{equation}
r=\frac{t}{V}[n_{s}\cos (\pi n_{s})-\frac{1}{\pi }\sin (\pi n_{s})],
\label{r}
\end{equation}
where $n_{s}=n/(1-n)$. The remaining terms of $H_{s}$, and the whole of $%
H_{d}$ can both be mapped into a spinless fermion model $H_{\alpha }^{sf}$ ($%
\alpha =s$ or $d$):

\begin{eqnarray}
H_{\alpha }^{sf} &=&\Delta _{\alpha }\sum_{i}f_{i}^{\dagger }f_{i}-t_{\alpha
}\sum_{i\sigma }(f_{i+1}^{\dagger }f_{i}+{\rm H.c.})  \nonumber \\
&&+V_{\alpha }f_{i}^{\dagger }f_{i}f_{i+1}^{\dagger }f_{i+1},  \label{hsf}
\end{eqnarray}
with $\Delta _{s}=0$. When $\alpha =s$, a site occupied by a fermion $%
f_{i}^{\dagger }$ corresponds to a single occupied site {\em and} an empty
site at the right of it, as explained in detail by Penc and Mila \cite{penc}
(a similar mapping was also used in a model for oxygen ordering in YBa$_{2}$%
Cu$_{3}$O$_{6+x}$ \cite{one}). Then, for $L$ sites and $N$ particles in $%
H_{s}$, the corresponding number of sites and particles in $H_{s}^{sf}$ are $%
L_{s}=L-N$, $N_{s}=N$. Then, the energy per site $e_{s}(n)$ of $H_{s}$ for
density $n$ is related to the corresponding quantity $e_{s}^{sf}(n_{s})$ of $%
H_{s}^{sf}$ by:

\begin{equation}
e_{s}(n)=\frac{E_{s}}{L}=\frac{L_{s}}{L}\frac{E_{s}^{sf}}{L_{s}}%
=(1-n)e_{s}^{sf}(n_{s}).  \label{es}
\end{equation}
Similarly, $H_{d}$ can be cast into the form of Eq. (\ref{hsf}), mapping a
doubly occupied site and an empty site at the right of it into a single site
occupied by a fermion. The mapping of the different physical quantities is
the same as that used before to find the correlation exponent $K_{\rho }$ in
a generalized $t-J$ model with very large three-site term \cite{bat,note}.
The number of sites and fermions in $H_{d}^{sf}$ are $L_{d}=L-N/2$, $%
N_{d}=N/2$. The energy per site of $H_{d}$ is given in terms of that of $%
H_{d}^{sf}$ by:

\begin{equation}
e_{d}(n)=(1-\frac{n}{2})e_{d}^{sf}(n_{d})\text{; }n_{d}=\frac{n}{2-n}.
\label{ed}
\end{equation}
To calculate $K_{\rho }$, we also need the mapping of the velocity \cite{bat}%
:

\begin{equation}
v_{d}=\frac{L}{L_{d}}v_{d}^{sf}=\frac{2}{2-n}v_{d}^{sf},  \label{v}
\end{equation}
and/or the Drude weight:

\begin{equation}
D_{d}=\frac{L}{2}\frac{\partial ^{2}E_{d}(\Phi )}{\partial \Phi ^{2}}=4\frac{%
L_{d}}{L}\frac{\partial ^{2}E_{d}^{sf}(\Phi _{sf})}{\partial \Phi _{sf}^{2}}%
=2(2-n)D_{d}^{sf}.  \label{d}
\end{equation}
Here $E_{d}(\Phi )$ ($E_{d}^{sf}(\Phi _{sf})$) is the energy of a ring
described by $H_{d}$ ($H_{d}^{sf}$) threaded by a flux $\Phi $ ($\Phi _{sf}$%
). The correlation exponent can be calculated as \cite{note}:

\begin{equation}
K_{\rho }=\frac{\pi v_{d}}{2\partial ^{2}e_{d}/\partial n^{2}}=\frac{\pi
D_{d}}{v_{d}}=\pi (\frac{D_{d}}{2\partial ^{2}e_{d}/\partial n^{2}})^{1/2}.
\label{k}
\end{equation}
Similar expressions give $K_{\rho }$ for $H_{s}$, but we do not give them,
since in the M phase always $K_{\rho }<1$, and we are interested in the
region $K_{\rho }>1$, for which superconducting correlations dominate at
large distances.

Using a Jordan-Wigner transformation, $H_{\alpha }^{sf}$ is transformed into
an equivalent $XXZ$ model with $L_{\alpha }$ sites and $M_{\alpha }$ spins
down:

\begin{eqnarray}
H_{\alpha }^{sf} &\equiv &H_{\alpha }^{XXZ}=2t_{\alpha
}\sum_{i}(S_{i}^{x}S_{i+1}^{x}+S_{i}^{y}S_{i+1}^{y})+V_{\alpha
}\sum_{i}S_{i}^{z}S_{i+1}^{z}  \nonumber \\
&&+\Delta _{\alpha }N_{\alpha }+V_{\alpha }(N_{\alpha }-L_{\alpha }/4).
\label{xxz}
\end{eqnarray}
We have calculated the energy $e_{\alpha }^{sf}(n_{\alpha })$ solving
numerically the integral equations of the exact Bethe ansatz solution of Eq.
(\ref{xxz}) in the thermodynamic limit \cite{yan}. To obtain $K_{\rho }$, we
have calculated the excitation energies for two small momenta, solving
numerically the corresponding Bethe ansatz equations \cite{fow}. This
allowed us to extract $v_{d}^{sf}(n_{d})$. From it, the numerical second
derivative of $e_{d}(n)$, Eqs. (\ref{ed}), (\ref{v}) and the first Eq. (\ref
{k}), $K_{\rho }$ was calculated for $n_{d}\neq 1/2$. For $n_{d}=1/2$, we
had technical problems in the calculation of $v_{d}^{sf}(n_{d})$, but
fortunately, analytical expressions are known \cite{sri,gia}:

\begin{eqnarray}
v_{d}^{sf} &=&\frac{\pi t\sin \mu }{\mu }\text{; }D_{d}^{sf}=\frac{v_{d}^{sf}%
}{4(\pi -\mu )};  \nonumber \\
\mu &=&\arccos (V/2t).  \label{ana}
\end{eqnarray}
Using Eqs. (\ref{v}), (\ref{d}), (\ref{k}) and (\ref{ana}), one has for $%
n=2/3$ ($n_{d}=1/2$):

\begin{equation}
K_{\rho }(2/3)=\frac{4}{9(1-\mu /\pi )}.  \label{k2s3}
\end{equation}
For $V_{\alpha }=0$, $H_{\alpha }^{sf}$ can be solved trivially \cite{bat}
and Eqs.(\ref{ed}), (\ref{v}), (\ref{d}), (\ref{k}) lead to another
analytical result for $V_{d}=0:$

\begin{equation}
K_{\rho }(2/3)=\frac{(2-n)^{2}}{2}.  \label{kv0}
\end{equation}
Since always $V_{d}/t_{d}>0$, and $K_{\rho }$ decreases with increasing $%
V_{d}$, Eq. (\ref{kv0}) implies that for large $V$, no phase with dominant
superconducting correlations exists for $n\geq 2-\sqrt{2}\simeq 0.59$.

\section{Bethe ansatz results}

\begin{figure}
\narrowtext
\epsfxsize=3.0truein
\vbox{\hskip 0.05truein \epsffile{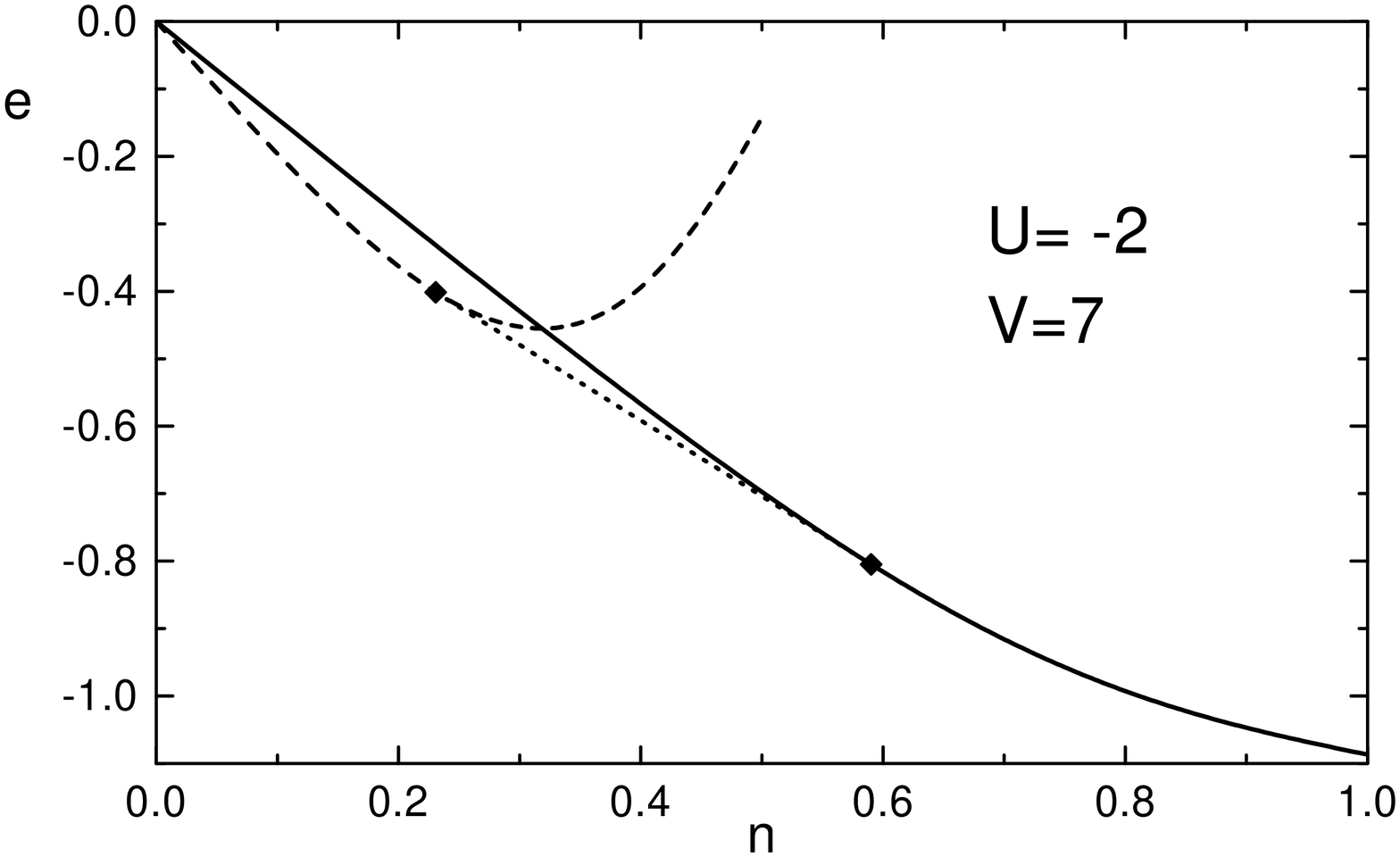}}
\medskip
\caption{Ground state energy of the phase with singly occupied sites
(dashed line) and doubly occupied sites (full line) as a function of
density. The dotted line is the Maxwell construction (see text).}
\label{fig1}
\end{figure}

In Fig. 1 we show the energy per site of $H_{s}$ ($e_{s}(n)$) with the last
term neglected, and that of $H_{d}$ ($e_{d}(n)$), as a function of density $%
n $ for $U=-2$ and $V=7$. We take $t=1$ as the unit of energy. The dotted
line represents the energy for average composition $n$, of an inhomogeneous
(phase separated) mixture of the ``metallic'' (M) phase described by $H_{s}$
for density $n_{1}=0.2307$ and the ``bipolaron'' (BP) phase (ground state of 
$H_{d}$) for density $n_{2}=0.5901$. These compositions $n_{i}$, represented
by diamonds in Fig. 1, are obtained finding the common tangent to both
curves $e_{s}(n)$ and $e_{d}(n)$ (Maxwell construction). Between them, the
energy of the phase separated phase is lower than both $e_{s}(n)$ and $%
e_{d}(n)$. The energy $e_{s}(n)$ is dominated by the first term of Eq. (\ref
{hs}), which already exists for $V=+\infty $. The effect of finite large $V$
is small (except near $n=1/2$ for which the first term of Eq. (\ref{hs})
vanishes), and does not change $n_{1}$ significantly. Instead, the effect of
a finite large $V$ on $n_{2}$ is dramatic, reducing it from $n_{2}=1$ to $%
n_{2}<0.6$. This is because for $V=+\infty $, $e_{d}(n)$ is always a
straight line (extending from $e_{d}(0)=0$ to $e_{d}(1)=-1$ if $U=-2$). The
second and third term of $H_{d}$ (Eq.(\ref{hd})), taken into account
exactly, are responsible of the curvature of $e_{d}(n)$ and the shift of $%
n_{2}$ from 1. This is the main effect of finite large $V$.

\begin{figure}
\narrowtext
\epsfxsize=3.0truein
\vbox{\hskip 0.05truein \epsffile{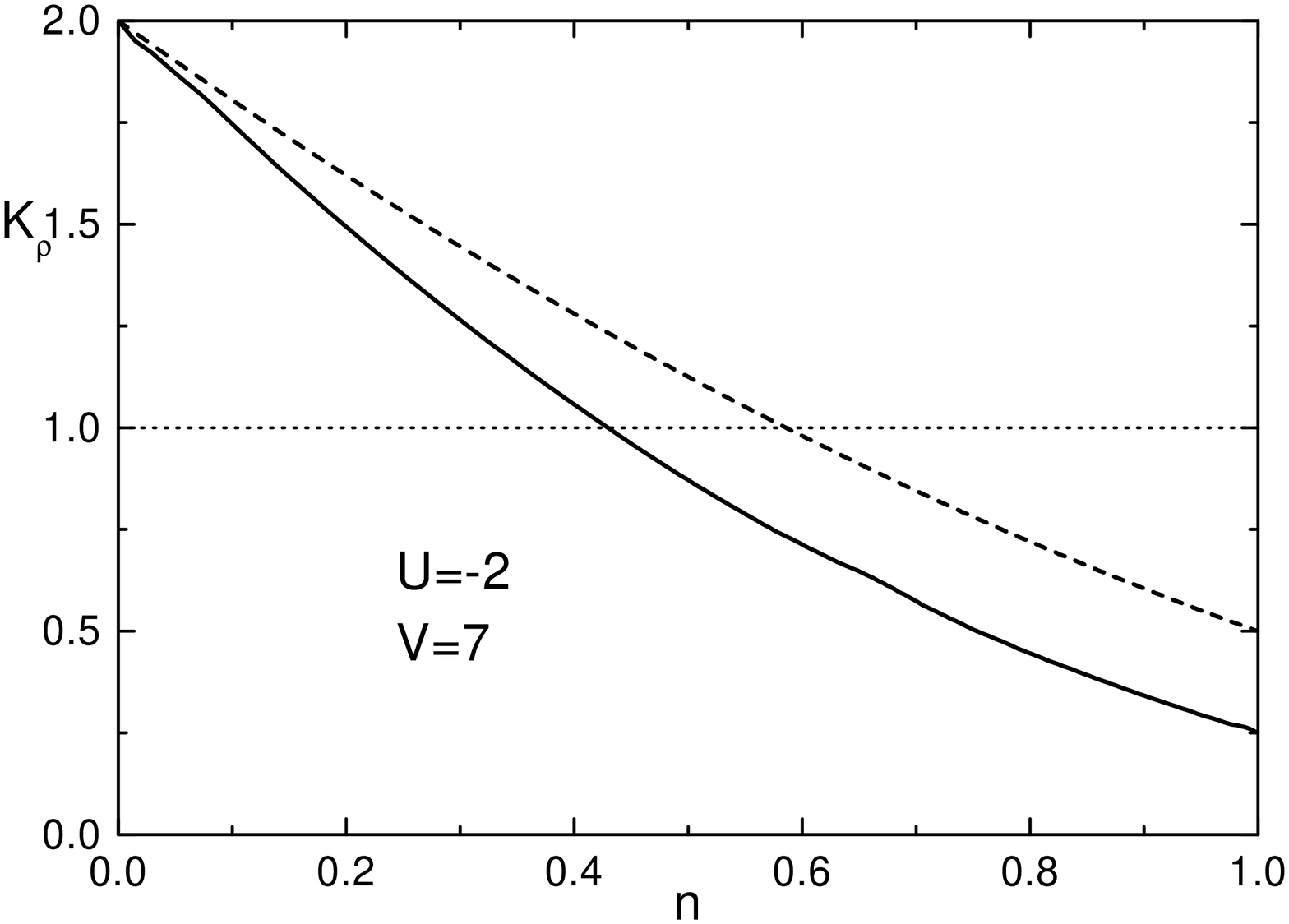}}
\medskip
\caption{Correlation exponent $K_{\rho }$ as a function of density in
the ``bipolaron'' (BP) phase for $U=-2$ and $V=7$ (full line). The dashed
line is the result in absence of the effective nearest-neighbor repulsion
between doubly occupied sites $(V_{d}=0)$.}
\label{fig2}
\end{figure}

Another important effect of a finite $V$ is that the pairs acquire mobility
and for low densities, superconducting correlations dominate at large
distances ($K_{\rho }>1$) in the BP phase. In Fig. 2 we represent $K_{\rho
}(n)$ in this phase for the same parameters of Fig. 1. Also shown is the
analytical result Eq. (\ref{kv0}), which is an upper bound of $K_{\rho }$
for any values of $U<0$ and $V>>t$. For $V\rightarrow +\infty $, we obtain a
critical density $n_{c}=0.422$ for which $K_{\rho }(n_{c})=1$. For $n<n_{c}$
and arbitrary values of $U<0$ and $V>>t$, the exponent $K_{\rho }(n)>1$.
Thus, $n_{c}$ lies in the interval $(0.422,2-\sqrt{2})$. For $-5\leq U\leq 0$
and $5\leq V\leq 20$, we find that $n_{c}<0.45$. Also, $n_{c}$ depends very
weakly on $U$ and $V$ within the studied range of parameters, decreasing
with increasing repulsions. For $U=-2$, $V=7$ (as in Figs. 1 and 2), $%
n_{c}=0.429$. These results disagree with those obtained by numerical
diagonalization of small systems \cite{penc,sano}, which obtained $K_{\rho
}>1$ for small values of $|U|$, large $V$ and $n=1/2$, $n=2/3$, but agree
with the statement that the system might not be a Luttinger liquid in that
region \cite{penc}, and with recent Monte Carlo results in larger systems 
\cite{clay}. These results and the ones shown below indicate that there is
phase separation (PS)\ in that region. We believe that the reason of the
artificially large values of $K_{\rho }$ in the above mentioned numerical
results is that they were calculated using the first Eq. (\ref{k}) with $%
\partial ^{2}e/\partial n^{2}=1/(\kappa n^{2})$ determined numerically from
the energy for $N$, $N-2$ and $N+2$ particles with $N/L=n$. In phase
separated regions (dashed line in Fig. 1), the compressibility $\kappa $
diverges in the thermodynamic limit, but in small systems, $\kappa $ can be
large and positive due to finite-size effects, leading to very large values
of $K_{\rho }$, while in fact the system is not a Luttinger liquid. This
effect was present in numerical studies of the one-dimensional $t-J$ model
with correlated hopping \cite{lem}.

\begin{figure}
\narrowtext
\epsfxsize=3.0truein
\vbox{\hskip 0.05truein \epsffile{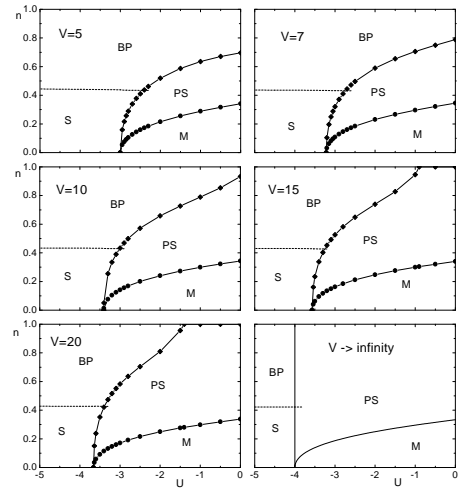}}
\medskip
\caption{Phase diagram of the model in the density-$U$ plane, for
different values of $V$. The phase without doubly occupied sites is called
``metallic'' (M), while the phase with no singly occupied sites is denoted
as ``bipolaronic'' (BP), and if $K_{\rho }>1$ we call it ``superconducting''
(S). The region of phase separation is labeled PS.}
\label{fig3}
\end{figure}

In Fig. 3 we show the phase diagram of the model, determined using the
Maxwell construction (as in Fig. 1) for different values of $U$ and $V$
(points indicated with solid symbols in Fig. 3), and searching the critical
densities $n_{c}$ for which $K_{\rho }(n_{c})=1$. The phase diagram for $%
V\rightarrow +\infty $ was already known, except for the boundary at $%
n=0.422 $, which separates the BP phase with $K_{\rho }<1$ (in which charge
correlation functions are the dominant ones at large distances) from the
``superconducting'' (S) phase with $K_{\rho }>1$. The most noticeable effect
of a finite $V$, already present for $V$ as large as $20t$ is the change in
the boundary between the BP and PS regions, as a consequence of the above
mentioned curvature of $e_{d}(n)$ for $V<+\infty $. Also, as $V$ decreases,
the region S with $K_{\rho }>1$ increases, moving to larger values of $U$
and to slightly larger densities. Finally, the PS region is reduced.

The case $V=5$ is probably beyond the quantitative validity of our large $V$
approximation. Monte Carlo results for $V=8$ (Fig. 14 of Ref. \cite{clay})
are in qualitative agreement with our results. Quantitatively, Clay {\it et
al.} obtain that for $U=0$, phase separation begins at $n_{1}\sim 0.5$,
while for $V=7$ we obtain $n_{1}=0.345$. This difference might be due to the
effect of terms of order $t^{m}$ with $m>2$ or the third term of $H_{s}$
(Eq. (\ref{hs})), which we have neglected. Terms of order $t^{4}$ reduce $%
t_{d}$ and increase $n_{1}$. The limit between S and M phases at low
densities is affected neither by $H_{sd}$ (Eq. (\ref{hsd})) nor by the third
term of $H_{s}$, but might be changed slightly by terms of order $t^{4}$.

\section{Summary and discussion}

Generalizing a previous approach \cite{penc}, we have constructed an
effective Hamiltonian $H_{eff}$ for the extended Hubbard model when $V>>t$
and $U\leq 0$. This permits to study the region $V>8t$, which is outside the
region of validity of weak-coupling approaches \cite{voitl,frad,voitb,bos},
while numerical diagonalization of small systems \cite{penc,sano,ali}
display important finite-size effects, as discussed in Section III, and
Monte Carlo calculations have some technical problems \cite{clay}. The
effective Hamiltonian can be divided in three parts: $H_{s}$, $H_{d}$ and $%
H_{sd}$. The latter is irrelevant for sufficiently large $V$, and $H_{s}$ ($%
H_{d}$) acts on a phase in which all particles move in singly (doubly)
occupied sites. $H_{d}$ and the most important terms of $H_{s}$ were mapped
into a Bethe ansatz exactly solvable model, which allows to obtain the
energy and correlation exponent $K_{\rho }$ in the thermodynamic limit. From
this information we have constructed the phase diagram. We obtain a region
for low and intermediate densities and sufficiently negative $U$, in which
the system behaves as a Luttinger liquid with dominant superconducting
correlations at large distances, for any finite $V>>t$. The existence of
this phase and the main changes in the phase diagram with respect to the $%
V=+\infty $ limit, are due to the dynamics of doubly occupied sites in $H_{d}
$, controled by terms of order $t^{2}/(V-U)$. The most noticeable effect of
a finite $V$ for moderate values of $|U|$ ($U>-4$), is that the upper
density of the phase separation (PS) region is reduced from $n=1$ to much
lower values. Instead, the larger density for which superconducting
correlations dominate at large distances remains near $n=0.42$. Thus, the
place left by the PS region is mostly occupied by the BP (``bipolaronic''
``normal'') phase rather than by the S (``superconducting '') phase.
Nevertheless, as $V$ decreases from very large values to $V\sim 7$, the
upper value of $U$ for which the S phase exists increases from -4 to $\sim -3
$.

The extended Hubbard model with attractive $U$ can be justified in different
ways, as a model for doped BaBiO$_{3}$, as discussed in Section I. In spite
of the different dimensionality of the real compound, one can discuss
qualitatively the physics expected from the phase diagram (Fig. 3). BaBiO$%
_{3}$ has one electron per site ($n=1$), and the ground state of the model
is a charge density wave, in which nearest-neighbor sites are not
equivalent, as experimentally observed \cite{cox,chai,pei}. For sufficiently
negative $U$, as $n$ decreases (corresponding to partial replacement of Ba
for K) keeping $n_{c}\sim 0.4<n<1$, the ground state of the model is a
Luttinger liquid with dominant charge density wave correlations at large
distances, with wave vector $2k_{F}=\pi n$. Experimentally, the ground state
of Ba$_{1-x}$K$_{x}$BiO$_{3}$ has different charge density wave orderings
for $0\leq x<\sim 0.4$ ($1\geq n>\sim 0.6$) \cite{pei}. As $n$ is further
lowered (corresponding to increasing $x=1-n$), the model enters a region
with dominant superconducting correlations at large distances, which has a
corresponding superconducting phase in Ba$_{1-x}$K$_{x}$BiO$_{3}$ (for $\sim
0.4<x<0.5$) \cite{pei}. Above $x=0.5$ the experimental system cannot be
formed, since the solubility limit of K atoms is exceeded. This picture is
also consistent with mean-field calculations in the three-dimensional model
for different parameters \cite{varm}.

\section*{Acknowledgments}

We are partially supported by CONICET. This work was sponsored by PICT
03-00121-02153 of ANPCyT and PIP 4952/96 of CONICET.


\begin{references}
\bibitem{slei}  A.W. Sleight, J.L. Gillson, and P.E. Bierstadt, Solid State
Commun. {\bf 17}, 27 (1975).

\bibitem{matt}  L.F. Mattheiss, E.M. Gyorgy, and D.W. Johnson, Jr., Phys.
Rev. B {\bf 37}, 3745 (1988); R.J. Cava, B. Battlog, J.J. Krajewski, R.
Farrow, L.W. Rupp Jr., A.E. White, K. Short, W.F. Peck, and T. Kometani,
Nature (London) {\bf 332}, 814 (1988).

\bibitem{batt}  B.Battlog, R.J. Cava, L.W. Rupp Jr., A.M. Mujsce, J.J.
Krajewski, J.P. Remeika, W.F. Peck Jr., A.S. Cooper, and G.P. Espinosa,
Phys. Rev. Lett {\bf 61}, 1670 (1988).

\bibitem{loon}  C.K. Loong, P. Vashishta, R.K. Kalia, M.H. Degani, D.L.
Price, J.D. Jorgensen, D.G. Hinks, B. Dabrowski, A.W. Mitchell, D.R.
Richards, and Y. Zheng, Phys. Rev. Lett. {\bf 62}, 2628 (1989).

\bibitem{cox}  D.E. Cox and A.W. Sleight, Acta Crystallogr. B {\bf 35}, 1
(1979).

\bibitem{chai}  C. Chaillout, A. Santoro, J.P. Remeika, A.S. Cooper, G.P.
Espinosa, and M. Marezio, Solid State Commun. {\bf 65}, 1363 (1988).

\bibitem{pei}  S. Pei, J.D. Jorgensen, B. Dabrowski, D.G. Hinks, D.R.
Richards, A.W. Mitchell, J.M. Newsam, S.K. Sinha, D. Vaknin, and A.J.
Jacobson, Phys. Rev. B {\bf 41}, 4126 (1990).

\bibitem{dolo}  M.D. N\'{u}\~{n}ez Regueiro and A.A. Aligia, Phys. Rev.
Lett. {\bf 61}, 1889 (1988); J. Bala and A.M. Ole\'{s}, Phys. Rev. B {\bf 47}%
, 515 (1993) ;A.A. Aligia, M.D. N\'{u}\~{n}ez Regueiro, and E.R. Gagliano, 
{\it ibid } {\bf 40}, 4405 (1989); A.A. Aligia and M. Bali\~{n}a, {\it ibid }%
{\bf 47}, 14380 (1993).

\bibitem{sofo}  J.O. Sofo, A.A. Aligia, and M.D. N\'{u}\~{n}ez Regueiro,
Phys. Rev. B {\bf 39}, 9701 (1989); {\bf 40}, 6955 (1989).

\bibitem{varm}  C.M. Varma, Phys. Rev. Lett. {\bf 61}, 2713 (1988).

\bibitem{nguy}  D. Nguyen Manh, D. Mayou and F. Cyrot-Lackmann, Solid State
Commun. {\bf 79}, 723 (1991).

\bibitem{rice}  T.M. Rice and L. Sneddon, Phys. Rev. Lett. {\bf 47}, 689
(1981).

\bibitem{micn}  R. Micnas, J. Ranninger and S. Robaszkiewicz, Rev. Mod.
Phys. {\bf 62}, 113 (1990); references therein.

\bibitem{sale}  S. Salem-Sugui, Jr., E.E. Alp, S.M. Mini, M. Ramanathan,
J.C. Campuzano, G. Jennings, M. Faiz, S. Pei, B. Dabrowski, Y. Zheng, D.R.
Richards, and D.G. Hinks, Phys. Rev. B {\bf 43}, 5511 (1991).

\bibitem{heal}  S.M. Heald, D. Di Marzio, M. Croft, M.S. Hedge, S. Li, and
M. Greenblatt, Phys. Rev. B {\bf 40}, 8828 (1989).

\bibitem{uchi}  S. Uchida, S. Tajima, A. Masaki, S. Sugai, K. Kitazawa, and
S. Tanaka, J. Phys. Soc. Jpn {\bf 54}, 4395 (1985).

\bibitem{sim}  M.E. Simon, A.A. Aligia, and E. Gagliano, Phys. Rev. B {\bf 56%
}, 5637 (1997); references therein. H. Rosner, H. Eschrig, R. Hayn, S.-L.
Drechsler, and J. M\'{a}lek, {\it ibid }{\bf 56}, 3402 (1997); references
therein..

\bibitem{mila}  F. Mila and X. Zotos, Europhys. Lett. {\bf 24}, 133 (1993).

\bibitem{penc}  K. Penc and F. Mila, Phys. Rev. B {\bf 49}, 9670 (1994).

\bibitem{sano}  K. Sano and Y. Ono, J. Phys. Soc. Jpn. {\bf 63}, 1250 (1994).

\bibitem{lin}  H.Q. Lin, E.R. Gagliano, D.K. Campbell, E.H. Fradkin, and
J.E. Gubernatis, in ``{\it The Physics and the Mathematical Physics of the
Hubbard Model '', }edited by D. Baeriswyl {\it et al. }(Plenum, New York,
1995).

\bibitem{ali}  A.A. Aligia, Europhys. Lett. {\bf 45}, 411 (1999).

\bibitem{clay}  R.T. Clay, A.W. Sandvik, and D.K. Campbell, Phys. Rev. B 
{\bf 59}, 4665 (1999).

\bibitem{nak}  M. Nakamura, cond-mat/9909277.

\bibitem{voitl}  J. Voit, Phys. Rev. Lett. {\bf 64} 323 (1990).

\bibitem{frad}  J.W. Cannon and E. Fradkin, Phys. Rev. B {\bf 41}, 9435
(1990).

\bibitem{voitb}  J. Voit, Phys. Rev. B {\bf 45}, 4027 (1992).

\bibitem{bos}  A.A. Aligia and L. Arrachea, Phys. Rev. B in press (BG7464).

\bibitem{one}  A.A. Aligia, Phys. Rev. B {\bf 47}, 15308 (1993).

\bibitem{bat}  C.D. Batista, F. Lema, and A.A. Aligia, Phys. Rev. B {\bf 52}%
, 6223 (1995).

\bibitem{note}  The correlation exponent $K_{\rho }$ refers to the original
model Eq. (\ref{ehm}), and not to the spinless models. In general, the
relations defining $K_{\rho }$ depend on the number of branches around the
Fermi level [see for example G. Santoro, N. Manini, A. Parola, and E.
Tosatti, Phys. Rev. B {\bf 53}, 828 (1996)].

\bibitem{yan}  C.N. Yang and C.P. Yang, Phys. Rev. {\bf 150}, 321 (1996); 
{\bf 150}, 327 (1996).

\bibitem{fow}  M. Fowler and M.W. Puga, Phys. Rev. B {\bf 18}, 421 (1978).

\bibitem{sri}  B. Sriram Shastry and B. Sutherland, Phys. Rev. Lett. {\bf 65}%
, 243 (1990).

\bibitem{gia}  T. Giamarchi and A.M. Tsvelik, Phys. Rev. B {\bf 59}, 11398
(1999); references therein.

\bibitem{lem}  F. Lema, C.D. Batista, and A.A. Aligia, Physica C {\bf 259},
287 (1996).
\end{references}
\end{document}